\documentclass[conference, a4paper]{IEEEtran}
\IEEEoverridecommandlockouts

\usepackage[dvipdfmx]{graphicx}
\usepackage{url}

\usepackage{multirow}

\usepackage{cite}
\usepackage{amsmath,amssymb,amsfonts}
\usepackage{algorithmic}
\usepackage{graphicx}
\usepackage{xcolor}
\def\BibTeX{{\rm B\kern-.05em{\sc i\kern-.025em b}\kern-.08em
 T\kern-.1667em\lower.7ex\hbox{E}\kern-.125emX}}
\begin{document}

\title{Do new investment strategies \\ take existing strategies' returns\\ -- An investigation into agent-based models --
\thanks{Note that the opinions contained herein are solely those of the authors and do not necessarily reflect those of SPARX Asset Management Co., Ltd.}
}

\author{\IEEEauthorblockN{Takanobu Mizuta}
\IEEEauthorblockA{\textit{SPARX Asset Management Co. Ltd.,} \\ Tokyo, Japan \\ https://orcid.org/0000-0003-4329-0645} }

\maketitle

\IEEEpubidadjcol

\begin{abstract}
Commodity trading advisors (CTAs), who mainly trade commodity futures, showed good returns in the 2000s. However, since the 2010's, they have not performed very well. One possible reason of this phenomenon is the emergence of short-term reversal traders (STRTs) who prey on CTAs for profit. In this study, I built an artificial market model by adding a CTA agent (CTAA) and STRT agent (STRTA) to a prior model and investigated whether emerging STRTAs led to a decrease in CTAA revenue to determine whether STRTs prey on CTAs for profit. To the contrary, my results showed that a CTAA and STRTA are more likely to trade and earn more when both exist. Therefore, it is possible that they have a mutually beneficial relationship.

\end{abstract}

\begin{IEEEkeywords}
CTA, Preying Investment Strategy, Agent-Based Model, Artificial Market Model
\end{IEEEkeywords}

\section{Introduction}
Commodity trading advisors (CTAs), who mainly trade commodity futures, showed good returns in the 2000s. However, since the 2010's, they have not performed very well\cite{Clenow2012}.

Clenow \cite{Clenow2012} reported the following points as reasons for this: (1) the yield of government bonds decreased, (2) most CTAs utilized similar trend-following strategies and tended to trade at the same time, leading to the increased number of CTAs further impacting market prices, (3) the variation of investors and investment strategies was diversified by alternative investments spreading, and (4) short-term reversal traders (STRTs) who preyed on CTAs for profit emerged.

In regard to Point (2), Nagao and Kobayashi \cite{nagao2016} reported that CTAs utilized similar and simple trend-following strategies to enable people to build an imitated fund. On the basis of Point (4), Raschke and Connors \cite{Raschke1996} proposed a short-term reversal strategy to prey on CTAs for profit.

However, the four points are not only inadequately founded but also contradict each other. Point (2) implies that CTAs did not perform well due to the number of CTAs increasing, but Point (3) implies that the number of CTAs has relatively decreased. Point (4) implies that STRTs utilize reversal strategies to trade on the opposite side and reduce market impacts by CTAs, who utilize trend-following strategies. Therefore, Point (4) contradicts Point (2).
When one investment strategy fails to work, new strategies, especially those that use new technologies or faster tradings, have tended to be blamed for a long time. Since the second half of the 20th century, this tendency has become stronger, and the media have often criticized new investment strategies using new technologies, electronic trades, program trades, algorithm trades, high-frequency trades, and AI traders. However, these critics seem to be missing the point. Therefore, arguments that new strategies are used to impede existing strategies, as mentioned in Point (4), should be discussed with particular suspicion.

Empirical studies cannot be conducted to investigate situations that have never occurred in actual financial markets, for example, comparing whether a certain investment strategy exists. As so many factors affect price formation in actual markets, empirical studies cannot be conducted to isolate the direct effect of whether a certain investment strategy exists. In contrast, an artificial market model, which is an agent-based model for a financial market,\footnote{Excellent reviews include \cite{lebaron2006agent, chen2009agent, 7850016, mizuta2016SSRNrev, mizuta2019arxiv, mizuta2022aruka}.} can isolate the pure contributions of an investment strategy. These are the advantages of artificial market simulation studies. Articles in both Nature \cite{farmer2009economy} and Science \cite{Battiston818} have argued that artificial market studies are expected to contribute to a greater understanding of actual markets. Bookstaber \cite{Bookstaber2017} also argued that an agent-based model can investigate  financial crises better than classical economics.

Many previous artificial market studies have contributed to explaining the nature of financial market phenomena such as bubbles and crashes. Recent artificial market studies have also contributed to discussions of appropriate financial regulations and rules \cite{mizuta2016SSRNrev, mizuta2019arxiv, mizuta2022aruka}. The JPX Working Paper series includes various studies that contributed to such discussions\footnote{https://www.jpx.co.jp/english/corporate/research-study/working-paper/index.html}.

In this study, I built an artificial market model by adding a CTA agent (CTAA) and STRT agent (STRTA) to the prior model of Mizuta et al.\cite{mizuta2016ISAFM} and investigated whether emerging STRTAs led to a decrease in CTAA revenue to determine whether STRTs prey on CTAs for profit, as stated in Point (4).

\section{Model}
The model of Chiarella and Iori \cite{chiarella2002simulation} is very simple but replicates long-term statistical characteristics observed in actual financial markets: a fat tail and volatility clustering. In contrast, that of Mizuta et al. \cite{mizuta2016ISAFM} replicates high-frequency micro structures, such as execution rates, cancel rates, and one-tick volatility, that cannot be replicated with the model of Chiarella and Iori \cite{chiarella2002simulation}. Only fundamental and technical analysis strategies that exist generally for any market at any time\footnote{Many empirical studies using questionnaires found these strategies to be the majority, which are comprehensively reviewed by Menkhoff and Taylor \cite{10.2307/27646888}. The empirical study using market data by Yamamoto\cite{yamamotopredictor} showed that investors are switching fundamental and technical analysis strategies.} are implemented into the agent model.

The simplicity of the model is very important for this study because unnecessary replication of macro phenomena leads to models that are overfitted and too complex. Such models prevent the understanding and discovery of mechanisms affecting price formation because of the increase in related factors. I explain the basic concept for constructing  my artificial market model in the review article \cite{mizuta2019arxiv, mizuta2022aruka} and the Appendix ``Basic Concept for Constructing Model.''

In this study, the artificial market model was built by adding a CTAA and STRTA to the prior model of Mizuta et al.\cite{mizuta2016ISAFM}.

This model contains one stock. The stock exchange uses a continuous double auction to determine the market price. In the auction mechanism, multiple buyers and sellers compete to buy and sell financial assets in the market, and transactions can occur at any time whenever an offer to buy and one to sell match. The minimum unit of price change is $\delta P$. The buy-order price is rounded  down to the nearest fraction, and the sell-order price is rounded up to the nearest fraction.

The model includes $n$ normal agents (NAs), one CTAA, and one STRTA. The agents always place an order for only one share. I implemented the variable ``tick time'' $t$, which increases by one when an NA orders.

\subsubsection{Normal agent (NA)}
To replicate the nature of price formation in actual financial markets, I introduced the NA to model a general investor. The number of NAs is $n$. NAs can short sell freely. The holding positions are not limited, so NAs can take an infinite number of shares for both long and short positions. First, at time $t=1$, NA No. $1$ places an order to buy or sell its risk asset; then, at $t=2,3,,,n$, NAs No. $2,3,,,n$ respectively place buy or sell orders. At $t=n+1$, the model returns to the first NA and repeats this cycle. An NA determines the order price and buys or sells using a combination of a fundamental and technical analysis strategies to form an expectation of a risk asset's return. The expected return of agent $j$ for each risk asset at $t$ is
\begin{equation}
r^{t}_{e,j} = (w_{1,j} \ln{\frac{P_f}{P^{t-1}}} + w_{2,j}\ln{\frac{P^{t-1}}{P^{t-\tau _ j-1}}}+w_{3,j} \epsilon ^t _j )/\Sigma_i^3 w_{i,j} \label{eq1}
\end{equation}
where $w_{i,j}$ is the weight of term $i$ for agent $j$ and is independently determined by random variables uniformly distributed on the interval $(0,w_{i,max})$ at the start of the simulation for each agent. $\ln$ is the natural logarithm. $P_f$ is a fundamental value and is a constant. $P^t$ is a mid-price (the average of the highest buy-order price and the lowest sell-order price) at $t$, and $\epsilon ^t _ j$ is determined by random variables from a normal distribution with average $0$ and variance $\sigma _ \epsilon$. $\tau_j$ is independently determined by random variables uniformly distributed on the interval $(1,\tau _{max})$ at the start of the simulation for each agent\footnote{When $t< \tau _ j$, the second term of Eq. (\ref{eq1}) is zero.}.

The first term in Eq. (\ref{eq1}) represents a fundamental strategy: the NA expects a positive return when the market price is lower than the fundamental value, and vice versa. The second term represents a technical analysis strategy using a historical return: the NA expects a positive return when the historical market return is positive, and vice versa. The third term represents noise.

After the expected return has been determined, the expected price is
\begin{equation}
P^t_{e,j}= P^t \exp{(r^t_{e,j})}.
\end{equation}

Order prices are scattered around the expected price $P^t_{e,j}$ to replicate many waiting limit orders. An order price $P^t_{o,j}$ is determined by random variables uniformly distributed on the interval $(P^t_{e,j}-P_d, P^t_{e,j}+P_d)$ where $P_d$ is a constant and means dispersion of the order prices scattered. Whether the agent buys or sells is determined by the magnitude relationship between $P^t_{e,j}$ and $P^t_{o,j}$:

when $P^t_{e,j}>P^t_{o,j}$, the NA places an order to buy one share, and

when $P^t_{e,j}<P^t_{o,j}$, the NA places an order to sell one share\footnote{When $t<t_c$, to generate enough waiting orders, the agent places an order to buy one share when $P_f>P^t_{o,j}$, or to sell one share when $P_f<P^t_{o,j}$. \label{ft01}}. The remaining order is canceled $t_c$ after the order time.

\subsubsection{CTA agent(CTAA) and STRT agent(STRTA)}
\begin{figure}[t]
\begin{center}
\includegraphics[scale=0.35]{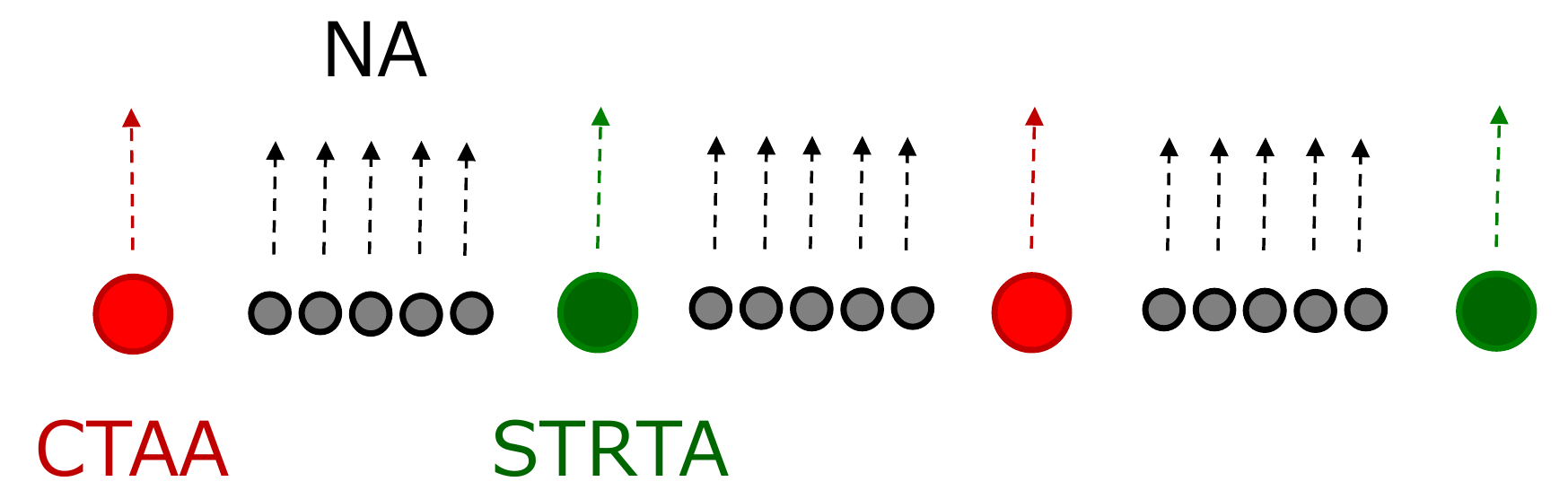}
\end{center}
\caption{Order cycle of CTAA and STRTA}
\label{p01}
\end{figure}

One CTAA and STRTA exist. They are limited to hold only one share or minus one share. The CTAA orders at $n \times \delta t$ ($n$ is a natural number and $\delta t$ is constant), and the STRTA orders at $n \times \delta t + \delta t /2$. They order only when market prices and their holdings meet the following conditions\footnote{When $t<t_c$, they do not place an order to generate a sufficient number of waiting orders.}(see also Fig. \ref{p01}). The tick time $t$ does not increase whether they order.

Trades of the CTAA are as follows. If the CTAA has no share, it places a market buy (sell) order\footnote{When an agent orders to buy (sell), if there is a lower sell-order price (a higher buy-order price) than the agent's order, dealing immediately occurs. Such an order is called a ``market order.''} of one share when $P^t$ reaches the maximum (minimum) within the last $\Delta T_1$ (and then it will have one (minus one) share). If the CTAA has one (minus one) share, it places a market sell (buy-back) order of one share when $P^t$ reaches the minimum (maximum) within the last $\Delta T_2$ (and then it will have no share).

Trades of the STRTA are as follows. If the STRTA has no share, it places a market buy (sell) order of one share when $P^t$ reaches the minimum (maximum) within the last $\Delta T_2$ (and then it will have one (minus one) share). If the STRTA has one (minus one) share, it places market sell (buy-back) order of one share when $P^t$ reaches the maximum (minimum) within the last $\Delta T_3$ (and then it will have no share).

\begin{figure}[t]
\begin{center}
\includegraphics[scale=0.35]{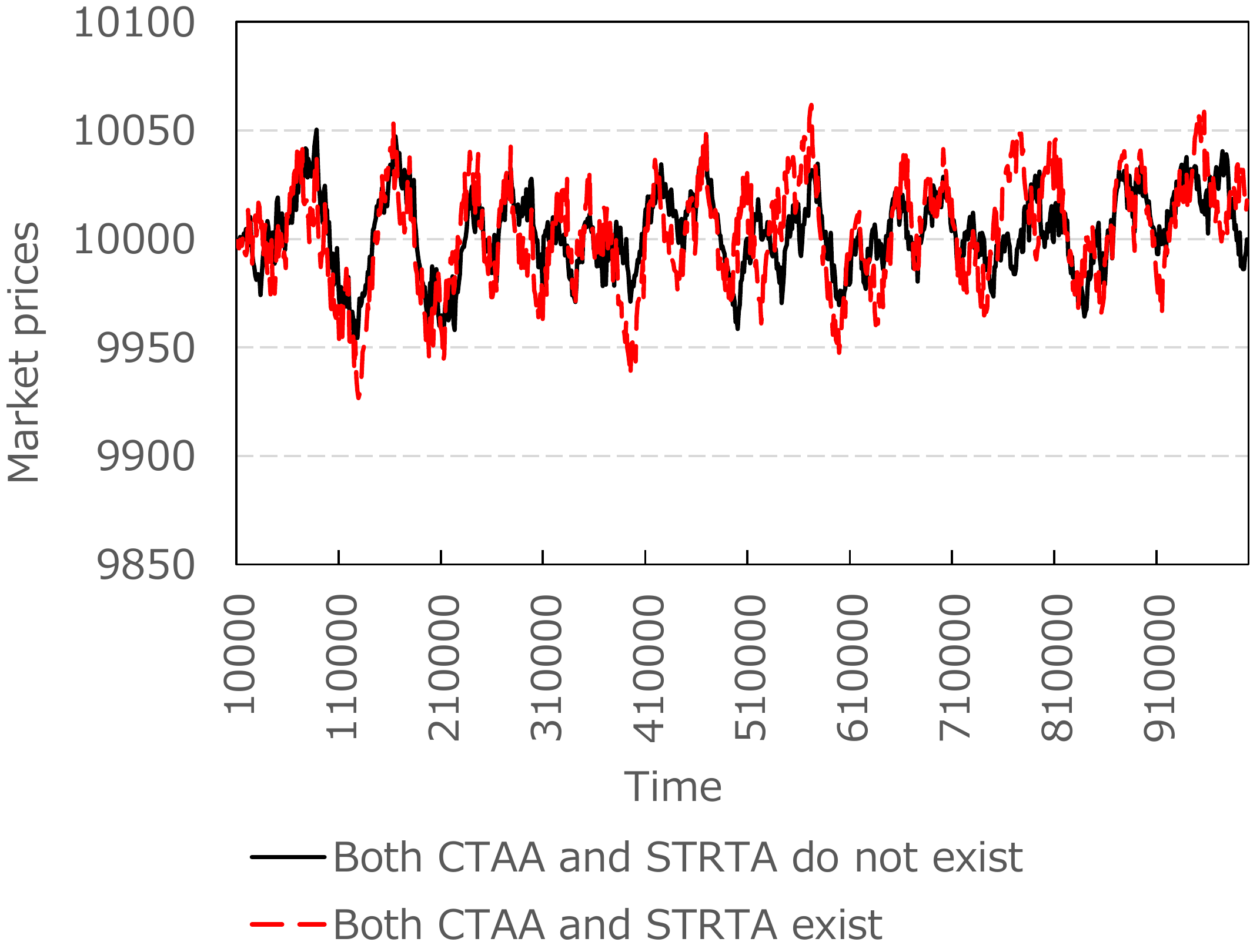}
\end{center}
\caption{Time evolution of prices when both CTAA and STRTA exist and not}
\label{z01}
\end{figure}

\begin{table*}[t]
\begin{center}
\begin{tabular}{cc}

\begin{minipage}{.30\textwidth}
\caption{Returns of CTAA}
\begin{center}
 \begin{tabular}{cc|rr}
\multicolumn{2}{c|}{} & \multicolumn{2}{c}{STRTA} \\
\multicolumn{2}{c|}{} & exist & not exist \\ \hline
\multirow{2}{*}{CTAA} & exist & 105\% & 90\% \\
 & not exist & 93\% & 79\% \\
\end{tabular}
\label{t01}
\end{center}
\end{minipage}

\begin{minipage}{.30\textwidth}
\caption{Trading volume of CTAA}
\begin{center}
 \begin{tabular}{cc|rr}
\multicolumn{2}{c|}{} & \multicolumn{2}{c}{STRTA} \\
\multicolumn{2}{c|}{} & exist & not exist \\ \hline
\multirow{2}{*}{CTAA} & exist & 892 & 865 \\
 & not exist & 891 & 866 \\
\end{tabular}
\label{t02}
\end{center}
\end{minipage}

\\

\begin{minipage}{.30\textwidth}
\caption{Returns of STRTA}
\begin{center}
 \begin{tabular}{cc|rr}
\multicolumn{2}{c|}{} & \multicolumn{2}{c}{STRTA} \\
\multicolumn{2}{c|}{} & exist & not exist \\ \hline
\multirow{2}{*}{CTAA} & exist & 293\% & 205\% \\
 & not exist & 265\% & 185\% \\
\end{tabular}
\label{t03}
\end{center}
\end{minipage}

\begin{minipage}{.30\textwidth}
\caption{Trading volume of STRTA}
\begin{center}
 \begin{tabular}{cc|rr}
\multicolumn{2}{c|}{} & \multicolumn{2}{c}{STRTA} \\
\multicolumn{2}{c|}{} & exist & not exist \\ \hline
\multirow{2}{*}{CTAA} & exist & 1739 & 1633 \\
 & not exist & 1685 & 1587 \\
\end{tabular}
\label{t04}
\end{center}
\end{minipage}

\\

\begin{minipage}{.30\textwidth}
\caption{Standard deviations of returns for $100$ tick times.}
\begin{center}
 \begin{tabular}{cc|rr}
\multicolumn{2}{c|}{} & \multicolumn{2}{c}{STRTA} \\
\multicolumn{2}{c|}{} & exist & not exist \\ \hline
\multirow{2}{*}{CTAA} & exist & 0.047\% & 0.038\% \\
 & not exist & 0.041\% & 0.032\% \\
\end{tabular}
\label{t05}
\end{center}
\end{minipage}

\begin{minipage}{.30\textwidth}
\caption{Standard deviations of returns for $20000$ tick times.}
\begin{center}
 \begin{tabular}{cc|rr}
\multicolumn{2}{c|}{} & \multicolumn{2}{c}{STRTA} \\
\multicolumn{2}{c|}{} & exist & not exist \\ \hline
\multirow{2}{*}{CTAA} & exist & 0.33\% & 0.28\% \\
 & not exist & 0.30\% & 0.25\% \\
\end{tabular}
\label{t06}
\end{center}
\end{minipage}

\end{tabular}
\end{center}
\end{table*}

\begin{figure}[t]
\begin{center}
\includegraphics[scale=0.35]{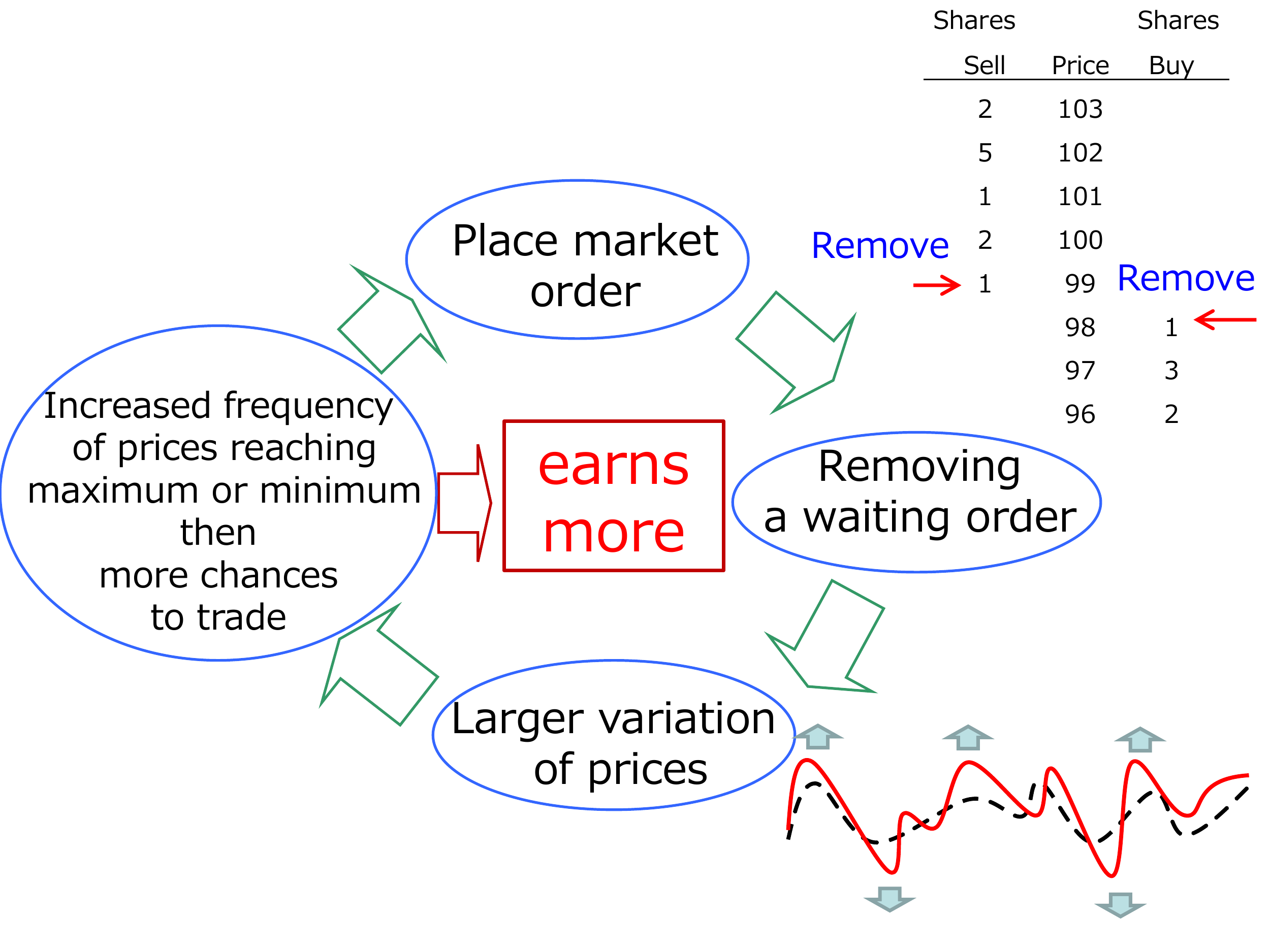}
\end{center}
\caption{Mechanism of co-prosperity}
\label{p02}
\end{figure}

\section{Simulation results}
In this study, I set the same parameters as in \cite{mizuta2016ISAFM}\footnote{I explain how they validated their model in the Appendix ``Validation of the Model.''}. Specifically, I set $n=1000, w_{1,max}=1, w_{2,max}=10, w_{3,max}=1, \tau _ {max}=10000, \sigma _ \epsilon = 0.03, P_d= 1000, t_c=10000, \delta P=0.01$, and $P_{f}=10000$. The simulations ran to $t=1000000$. I simulated four cases where the CTAA and STRTA existed or not, respectively, where the other parameters were fixed and used the same random number table. I then simulated these runs 100 times, changing the random number table each time, and reported the results as the averaged statistical values.

Fig. \ref{z01} shows the time evolution of the prices when both the CTAA and STRTA exist and when both do not exist. The two cases are not very different, and therefore trades of them do not impact the prices very much.

Tables \ref{t01} and \ref{t02} show the returns and trading volume of the CTAA for the four cases, respectively, where returns are defined as final earnings per the fundamental price ($P_f$). The results for when the CTAA does not exist are calculated on the basis that the CTAA would virtually trade with the best buy or sell price of waiting orders, and those orders would not change. As the CTAA does not exist, it cannot change the prices nor affect other agents, but the figures can be estimated.

In Table \ref{t01}, when the CTAA exists, it earns more when the STRTA exists than when it does not even thought the time evolution of prices are not so different as Fig. \ref{z01}. Table \ref{t02} shows that the CTAA can trade more if the STRTA exists than without it. This means that the CTAA has more chances of trades and can earn more when the STRTA exists.

Tables \ref{t03} and \ref{t04} show the returns and trading volume of the STRTA for the four cases, respectively. The STRTA trades and earns more when the CTAA exists similarly to the case of the CTAA in Tables \ref{t01} and \ref{t02}. In summary, the CTAA and STRTA have more chances of trades and earning more when both exist. Therefore, they have a mutually beneficial relationship.

Tables \ref{t05} and \ref{t06} show the standard deviations of returns for 100 (Table \ref{t05}) and 20000 (Table \ref{t06}) tick times for the four cases, respectively. The standard deviations are larger when both the CTAA and STRTA exist. The larger variation of prices leads to an increase in frequency that prices reach the maximum or minimum and that the CTAA and STRTA have more chances to trade.

Fig. \ref{p02} shows the mechanism derived from the aforementioned discussion. The CTAA and STRTA place a market order, which leads to a waiting order being removed from the order book. Decreasing waiting orders leads to other buy (sell) market orders executing with a waiting sell (buy) order of a higher (lower) price. This leads to a larger variation of prices and an increase in frequency that prices reach the maximum or minimum. Furthermore, the CTAA and STRTA have more chances to trade and earn more when each other exist, and increasing their trades leads to more market orders and the strengthening of this cycle.

\section{Summary}
In this study, I built an artificial market model by adding a CTAA and STRTA to the prior model of Mizuta et al.\cite{mizuta2016ISAFM} and investigated whether emerging STRTAs led to a decrease in CTAA revenue to determine whether STRTs prey on CTAs for profit.

To the contrary, my results showed that CTAAs and STRTAs have more chances of trade and earn more when each other exist. Therefore, they have a mutually beneficial relationship while some people says STRTs prey on CTAs for profit.

The mechanism of mutually beneficial relationship is following. The CTAA and STRTA place a market order, which leads to a waiting order being removed from the order book. Decreasing waiting orders leads to other buy (sell) market orders executing with a waiting sell (buy) order of a higher (lower) price. This leads to a larger variation of prices and an increase in frequency that prices reach the maximum or minimum. Furthermore, the CTAA and STRTA have more chances to trade and earn more when each other exist, and increasing their trades leads to make more market orders and the strengthening of this cycle.

When one investment strategy fails to work, new strategies, especially those that use new technologies or faster tradings, have tended to be blamed for a long time. Therefore, such arguments that new strategies are used to impede existing strategies should be discussed with particular suspicion. Since the second half of the 20th century, this tendency has become stronger, and the media have often criticized new investment strategies using new technologies, electronic trades, program trades, algorithm trades, high-frequency trades, and AI traders. However, these critics seem to be missing the point. These criticisms should be investigated using an artificial market model, which are considered future works.

\begin{table*}[t]
\caption{Stylized facts}
\begin{center}
 \begin{tabular}{ccrrrr}
 & CTAA & \multicolumn{2}{c}{exist} & \multicolumn{2}{c}{not exist} \\
 & STRTA & exist & not exist & exist & not exist \\ \hline
 \multicolumn{2}{c}{kurtosis or returns} &$3.46$ & $3.80$ & $3.67$ & $3.99$\\ \hline
 & lag & \\
 & 1 & $0.131$& $0.127$& $0.138$& $0.128$ \\
 autocorrelation & 2 & $0.081$& $0.084$& $0.078$& $0.080$ \\
 coefficient for & 3 & $0.061$& $0.069$& $0.059$& $0.060$ \\
 square returns & 4 & $0.054$& $0.058$& $0.047$& $0.051$ \\
 & 5 & $0.046$& $0.053$& $0.042$& $0.043$
 \end{tabular}
\end{center}
\label{t0}
\end{table*}

\section*{Appendix}

\subsection{Basic concept for constructing a model}
An artificial market model, which is a kind of agent-based model, can be used to investigate situations that have never occurred, handle regulation changes that have never been made, and isolate the pure contribution of these changes to price formation and liquidity \cite{mizuta2019arxiv, mizuta2022aruka}. These are the  advantages of an artificial market simulation.

However, the outputs of this simulation would not be accurate or be credible forecasts of the actual future. The simulation needs to reveal possible mechanisms that affect price formation through many simulation runs, e.g., searching for parameters or purely comparing the before and after states of changes. The possible mechanisms revealed by these runs provide new intelligence and insight into the effects of the changes on price formation in actual financial markets. Other methods of study, e.g., empirical studies, would not reveal such possible mechanisms.

Artificial markets should replicate the macro phenomena that exist generally for any asset at any time. Price variation, which is a kind of macro phenomenon, is not explicitly modeled in artificial markets. Only micro processes, agents (general investors), and price determination mechanisms (financial exchanges) are explicitly modeled. Macro phenomena emerge as the outcome interactions of micro processes. Therefore, the simulation outputs should replicate existing macro phenomena to generally prove that simulation models are probable in actual markets.

However, it is not the primary purpose for an artificial market to replicate specific macro phenomena only for a specific asset or a specific period. Unnecessary replication of macro phenomena leads to models that are overfitted and too complex. Such models would prevent us from understanding and discovering mechanisms that affect price formation because the number of related factors would increase.

In addition, artificial market models that are too complex are often criticized because they are very difficult to evaluate\cite{chen2009agent}. A model that is too complex not only would prevent us from understanding mechanisms but also could output arbitrary results by overfitting too many parameters. It is more difficult for simpler models to obtain arbitrary results, and these models are easier to evaluate.

Therefore, I constructed an artificial market model that is as simple as possible and does not intentionally implement agents to cover all the investors who would exist in actual financial markets.

As Michael Weisberg mentioned\cite{Weisberg2012}, ``Modeling, (is) the indirect study of real-world systems via the construction and analysis of models.'' ``Modeling is not always aimed at purely veridical representation. Rather, they worked hard to identify the features of these systems that were most salient to their investigations.''

Therefore, effective models are different depending on the phenomena they focus on. Thus, my model is effective only for the purpose of this study and not for others. The aim of my study is to understand how important properties (behaviors, algorithms) affect macro phenomena and play a role in the financial system rather than representing actual financial markets precisely.

As Weisberg also mentioned\cite{Weisberg2012}, ``When one invokes a computational model to explain some phenomenon, one is typically using transition rules or algorithm as the explanans. Schelling explained segregation by pointing out that small decisions reflecting small amounts of bias will aggregate to massively segregated demographics\cite{Schelling2006}. Neither the time sequence of the model's states nor the final, equilibrium state of the model carries the explanatory force; the algorithm itself is needed.''

An aim of my study is to understand how important properties (behaviors, algorithms) affect the investigation of macro phenomena and play a role in the financial system rather than representing actual financial markets precisely.

The above discussion holds not only for artificial markets but also for agent-based models used in other fields besides financial markets. For example, Thomas Schelling, who was received the Nobel prize in economics, used an agent-based model to discuss the mechanism of racial segregation. The model was built very simply compared with an actual town in order to focus on the mechanism\cite{Schelling2006}. Indeed, while it was not able to predict the segregation situation in the actual town, it was able to explain the mechanism of segregation as a phenomenon.

Harry Stevens, a newspaper writer, simulated an agent-based model to explain the spread of COVID-19 and to find a way how to prevent infection \cite{Stevens2020}. The model is too simple to replicate the real world, but its simplicity enabled it to reveal the mechanism behind the spread. 

Michael Weisberg studied what mathematical and simulation models are in the first place and cited the example of a map\cite{Weisberg2012}. Needless to say, a map models geographical features on the way to a destination. With a simple map we can easily understand the way to the destination. On the other hand, while a satellite photo replicates actual geographical features very well, we can not easily find the way to the destination. 

The title page of the book \cite{Weisberg2012} cited a passage from a short story by Jorge Borges\cite{Borges}, ``In time, those Unconscionable Maps no longer satisfied, and the Cartographers Guilds struck a Map of the Empire whose size was that of the Empire, and which coincided point for point with it . . . In the Deserts of the West, still today, there are Tattered Ruins of that Map, inhabited by Animals and Beggars.'' The story in which a map was enlarged to the same size as the real Empire to become the most detailed of any map is an analogy to that too detailed a model is not useful. This story give us one of the most important lessons for when we build and use any model.

\subsection{Validation of the model}
In many previous artificial market studies, the models were validated to determine whether they could explain stylized facts, such as a fat tail or volatility clustering \cite{lebaron2006agent,chen2009agent, mizuta2019arxiv, mizuta2022aruka}. A fat tail means that the kurtosis of price returns is positive. Volatility clustering means that square returns have a positive autocorrelation, and this autocorrelation slowly decays as its lag becomes longer.

Many empirical studies, e.g., that of Sewell \cite{Sewell2006}, have shown that both stylized facts (fat tail and volatility clustering) exist statistically in almost all financial markets. Conversely, they also have shown that only the fat tail and volatility clustering are stably observed for any asset and in any period because financial markets are generally unstable. This leads to the conclusion that an artificial market should replicate macro phenomena that exist generally for any asset at any time, fat tails, and volatility clustering. This is an example of how empirical studies can help an artificial market model.

The kurtosis of price returns and the autocorrelation of square returns are stably and significantly positive, but the magnitudes of these values are unstable and very different depending on the asset and/or period. Very broad magnitudes of about $1 \sim 100$ and about $0 \sim 0.2$, respectively, have been observed \cite{Sewell2006}.

For the aforementioned reasons, an artificial market model should replicate these values as significantly positive and within a reasonable range. It is not essential for the model to replicate specific values of stylized facts because the values of these facts are unstable in actual financial markets.

Table \ref{t0} lists the statistics showing the stylized facts, kurtosis of price returns for $100$ tick times, and autocorrelation coefficient for square returns for $100$ tick times for the four cases of whether CTAAs and STRTAs exist. This shows that this model replicated the statistical characteristics, fat-tails, and volatility clustering observed in real financial markets.

\bibliographystyle{IEEEtran}
\bibliography{ref}

\end{document}